\documentclass[preprint]{acmsiggraph}

\def\BibTeX{{\rm B\kern-.05em{\sc i\kern-.025em b}\kern-.08em
    T\kern-.1667em\lower.7ex\hbox{E}\kern-.125emX}}

\TOGonlineid{0248}

\TOGvolume{0}
\TOGnumber{0}

\usepackage{graphicx}
\graphicspath{{./figs/}}

\usepackage{wrapfig}

\usepackage[usenames,dvipsnames,svgnames,table]{xcolor}

\newif\ifsubmit
\submitfalse
\ifsubmit
    \newcommand{\todo}[1]{}
    \newcommand{\TODO}[1]{}
    
    \newcommand{\yotam}[1]{}
    \newcommand{\jm}[1]{}
    \newcommand{\jianchao}[1]{}
    
    \newcommand{\topic}[1]{}
    
    \newcommand{\change}[1]{}
\else
    \newcommand{\todo}[1]{\texttt{\textcolor{red}{TODO: #1}}}
    \newcommand{\TODO}[1]{\textcolor{red}{\textbf{****** #1 ******}}}

    \newcommand{\yotam}[1]{{\color{purple}\textsc{Yotam:} #1}}
    \newcommand{\jm}[1]{{\color{green}\textsc{Jyh-Ming:} #1}}
    \newcommand{\jianchao}[1]{{\color{orange}\textsc{Jianchao:} #1}}
    
    \newcommand{\topic}[1]{{\color{magenta}\textsc{Topic:} #1}}
    
    \newcommand{\change}[1]{\textcolor{blue}{#1}}
\fi

\usepackage{letltxmacro}
\LetLtxMacro{\oldmarginpar}{\marginpar}
\renewcommand{\marginpar}[1]{\oldmarginpar{\footnotesize #1}}

\usepackage{booktabs}
\usepackage{microtype}

\usepackage{mathtools}
\usepackage{amssymb}

\newcommand{\p}{\mathbf{p}}
\renewcommand{\c}{\mathbf{c}}

\newcommand*\patchAmsMathEnvironmentForLineno[1]{%
\expandafter\let\csname old#1\expandafter\endcsname\csname #1\endcsname
\expandafter\let\csname oldend#1\expandafter\endcsname\csname end#1\endcsname
\renewenvironment{#1}%
{\linenomath\csname old#1\endcsname}%
{\csname oldend#1\endcsname\endlinenomath}}%
\newcommand*\patchBothAmsMathEnvironmentsForLineno[1]{%
\patchAmsMathEnvironmentForLineno{#1}%
\patchAmsMathEnvironmentForLineno{#1*}}%
\AtBeginDocument{%
\patchBothAmsMathEnvironmentsForLineno{equation}%
\patchBothAmsMathEnvironmentsForLineno{align}%
\patchBothAmsMathEnvironmentsForLineno{flalign}%
\patchBothAmsMathEnvironmentsForLineno{alignat}%
\patchBothAmsMathEnvironmentsForLineno{gather}%
\patchBothAmsMathEnvironmentsForLineno{multline}%
\patchBothAmsMathEnvironmentsForLineno{eqnarray}%
}

\title{Decomposing Digital Paintings into Layers via RGB-space Geometry}

\author{Jianchao Tan\thanks{e-mail:jtan8@gmu.edu}\\George Mason University \and
Jyh-Ming Lien\thanks{jmlien@gmu.edu}\\George Mason University \and
Yotam Gingold\thanks{ygingold@gmu.edu}\\George Mason University}
\pdfauthor{Jianchao Tan, Jyh-Ming Lien, Yotam Gingold}


\begin{document}

\teaser{
\includegraphics[width=\textwidth]{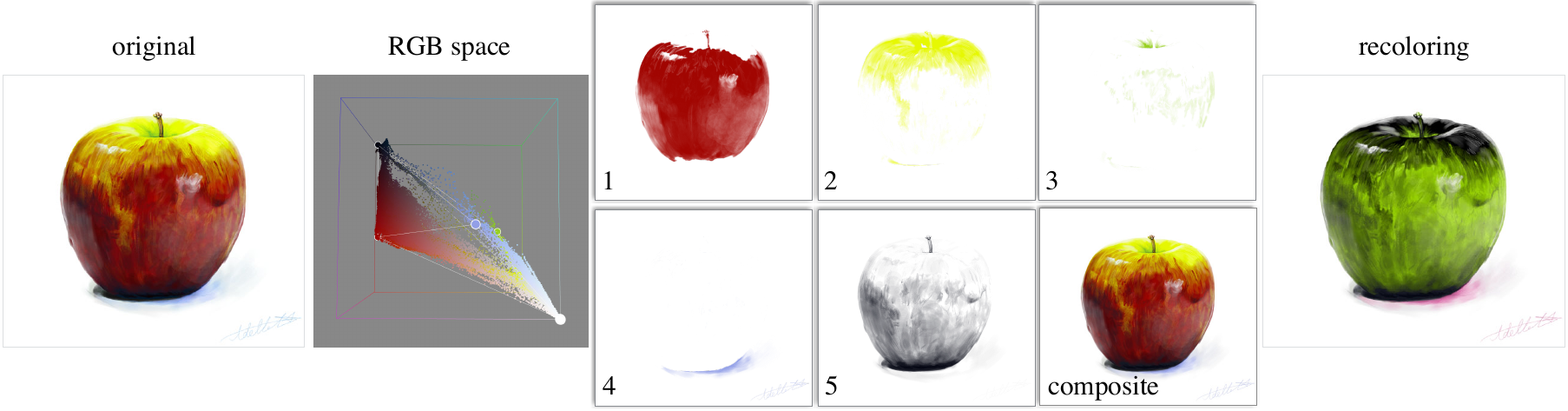}
\caption{Given a digital painting, we analyze the geometry of its pixels in RGB-space, resulting in a translucent layer decomposition, which makes difficult edits simple to perform.
Artwork \copyright\ 
Adelle Chudleigh.}
\label{fig:teaser}
}

\maketitle

\begin{abstract}

In digital painting software,
layers organize paintings.
However, layers are not explicitly represented, transmitted, or published
with the final digital painting.
We propose a technique to decompose a
digital painting into layers.
In our decomposition, each layer represents a coat of paint of
a single paint color applied
with varying opacity throughout the image.
Our decomposition is based on the painting's RGB-space geometry.
In RGB-space, a geometric structure is revealed
due to the linear nature of the
standard Porter-Duff~\shortcite{Porter:1984:CDI}
``over'' pixel compositing operation.
The vertices of the convex hull of
pixels in RGB-space suggest paint colors.
Users choose the degree of simplification
to perform on the convex hull, as well as a layer
order for the colors.
We solve a constrained optimization problem
to find maximally translucent, spatially coherent
opacity for each layer, such that the composition
of the layers reproduces the original image.
We demonstrate the utility of the
resulting decompositions for re-editing.

\end{abstract}

\begin{CRcatlist}
  \CRcat{I.3.7}{Computer Graphics}{Picture/Image Generation}{Bitmap and framebuffer operations}
  \CRcat{I.4.6}{Image Processing and Computer Vision}{Segmentation}{Pixel classification};
\end{CRcatlist}

\keywordlist

\copyrightspace

\section{Introduction}
\label{sec:intro}

Digital painting software simulates
the act of painting in the real world.
Artists choose paint colors and apply them
with a mouse or drawing tablet by
painting with a virtual brush.
These virtual brushes have varying opacity profiles,
which control how
the paint color blends with the background.
Digital painting software typically
provides the ability to create layers,
which are composited to form the final image
yet can be edited separately.
Layers \emph{organize} paintings.
However, layers are not explicitly represented, transmitted, or published
with the final digital painting.
Without layers, simple edits become extremely challenging, such as altering the color
of a coat without inadvertently affecting a scarf placed overtop.
Moreover, layers require artists to remember to create or switch layers when editing different parts of the painting.

We propose a technique to decompose a
digital painting into layers.
In our decomposition, each layer represents a coat of paint of
a single paint color applied
with varying opacity throughout the image.
We use the standard Porter-Duff~\shortcite{Porter:1984:CDI}
``$A$ over $B$'' compositing operation:
\begin{align}
\label{eq:porterduff}
	A_\alpha A_{RGB} + (1-A_\alpha)B_{RGB}
\end{align}
where the translucent pixel $A$ is placed over the pixel $B$ with translucency $\alpha \in [0,1]$.
Our technique automatically extracts paint colors,
and supports user intervention to choose a simplified palette with fewer colors.
Given a user-provided \emph{ordering} of the colors,
our technique computes per-pixel opacity values.\footnote{The ordering is necessary, because the ``over'' compositing operation is not commutative.}
The result is a sequence of layers that
reproduce the original painting when composited.
In RGB-space, the pixels of a digital painting
reveal a hidden geometric structure
(Figure~\ref{fig:rgbspace}).
This geometric structure
offers clues about the painting's editing history.
This structure results from
the linearity of the ``over'' compositing operation (Equation~\ref{eq:porterduff}),
which is the de facto standard for applying digital paint with transparency.
In this model, the paint color acts as a linear attractor
in RGB-space.
Affected pixels move towards the paint color via linear interpolation;
the painted stroke's transparency determines the strength of attraction,
or interpolation parameter.

Our \textbf{contributions} are:
\begin{itemize}
    \item The geometric analysis of RGB-space to determine the colors of paint used to create a digital painting (Section~\ref{sec:colors}).
    We make use of the RGB-space convex hull to obtain a small color palette capable of reproducing all others.
    \item An optimization-based approach to compute per-layer, per-pixel opacity values (Section~\ref{sec:opacity}).
    Our approach regularizes the original, underconstrained problem with terms that balance translucency and spatially coherence.
\end{itemize}
The result of these contributions is a technique that decomposes a single image, a digital painting, into translucent layers.
Our decomposition enables the structured re-editing of digital paintings.
\section{Related Work}
\label{sec:related}

\paragraph{Single-Image Decomposition}
Richardt et al.~\shortcite{Richardt:2014:VBS}
investigated a similar problem with the
goal of producing editable vector graphics.
Our goal is to produce editable layered bitmaps.
They proposed an approach in which
the user selects an image region,
and the region is then decomposed into
a linear or radial gradient and
the residual, background pixels.
Our approach targets bitmap graphics,
which are a less constrained domain.
Our approach also requires much less user input.
For comparison, we decompose an image from their paper
(the light in Figure~\ref{fig:gallery}, row 6).

Xu et al.~\shortcite{Xu:2006:ACP} presented an algorithm for
decomposing a single image of a Chinese painting into
a collection of layered brush strokes. Their 
approach is tailored to a particular style of artwork.
They recover painted colors by segmenting and fitting curves to brush strokes.
They also consider the problem of recovering
per-pixel opacity as we do. In their setting, however, they assume
at most two overlapping strokes and minimally varying transparency.
We consider a more general problem in which strokes have no known shape
and more than two strokes may overlap at once.
Fu et al.~\shortcite{Fu:2011:ACL} introduced a technique to
determine a plausible animated stroke order from a monochrome line drawing.
Their approach is based on cognitive principles, and operates on vector graphics.
We do not determine a stroke order; rather, we extract paint colors
and per-pixel opacity and operate on raster digital paintings.

McCann and Pollard~\shortcite{McCann:2009:LL,McCann:2012:SS} introduced
two generalizations to layering, allowing (a) pixels to have independent layer orders and (b) layers to partially overlap each other.
We solve for layer opacity coefficients in the traditional, globally and discretely ordered model of layers.

Scale-space filtering \cite{Witkin:1983:SSF} and related techniques \cite{Subr:2009:EMI,aujol2006color,farbman2008edge}
decompose a single image into levels of varying smoothness.
These decompositions separate the image according to
levels of detail, such as high frequency texture and
underlying low-frequency base colors.
These techniques are orthogonal to ours,
as they are concerned with spatial frequency
and we are concerned with color composition.

Intrinsic image decomposition
\cite{Grosse:2009:GTD,shen2008intrinsic,Bousseau:2009:UII}
attempts to separate a photographed object's illumination (shading) and reflectance (albedo).
This decomposition is suitable for photographs
of illuminated objects, but not digital paintings.

The recoloring approach of Chang et al.~\shortcite{Chang:2015:PPR}
extracts a color palette from a photograph by clustering.
Gerstner et al.~\shortcite{Gerstner:2013:PIA}
extracted sparse palettes from arbitrary images
for the purpose of creating pixel art.
Unlike approaches based on clustering the observed colors,
our approach has the potential to find simpler and even ``hidden'' colors.
Consider an image created from a blend of two colors with varying translucency, never opaque.
In the final image, the original colors will never be present, though an entire spectrum of other colors will be.

\paragraph{Editing History}
Tan et al.~\shortcite{Tan:2015:DTL} and Amati and Brostow~\shortcite{Amati:2010:MPI}
described approaches for decomposing time-lapse videos of
physical (and digital, for Tan et al.~\shortcite{Tan:2015:DTL})
paintings into layers. In our scenario, we have only the final
painting, though we make the simplifying assumption that
only Porter-Duff~\shortcite{Porter:1984:CDI} ``over'' blending operations were performed.

Hu et al.~\shortcite{Hu:2013:IIE} studied the problem
of reverse-engineering the image editing operation that occurred
between a pair of images.
We are similarly motivated by ``inverse image editing'',
though we solve an orthogonal problem in which only a single image is provided
and the only allowable operation is painting.

A variety of approaches have been proposed to make
use of image editing history (see Nancel and Cockburn~\shortcite{Nancel:2014:CCM} for a recent survey).
While we do not claim that our decomposition matches the
true image editing history, our approach could be used to provide
a plausible editing history. In particular, Wetpaint~\cite{Bonanni:2009:WST}
proposed a tangible ``scraping'' interaction for visualizing the layers of a painting.

\paragraph{Matting and Reflections}
Smith and Blinn~\shortcite{Smith:1996:BSM} studied
the problem of separating
a potentially translucent foreground object
from known backgrounds in a photo or video
(``blue screen matting'').
Zongker et al.~\shortcite{Zongker:1999:EMC} solved a
general version of this problem which allows for reflections and refractions.
Levin et al.~\cite{levin2008closed,levin2008spectral} presented solutions to the natural image matting problem,
which decomposes a photograph with a natural background into layers;
Levin et al.'s solutions assume as-binary-as-possible opacity values and at most three layers present in a small image patch.
Layer extraction has been studied in the context of photographs of
reflecting objects, such as windows~\cite{Szeliski:2000:LEM,farid1999separating,Levin:2004:SRS,sarel2004separating}.
These approaches make physical assumptions about the scene in the photograph,
they require a pair of photographs as input (\cite{farid1999separating}).
We consider digital paintings, in which physical assumptions are not valid
and there are typically more than two layers.

\section{Overview}
\label{sec:overview}

\begin{figure}
\centering
\raisebox{-0.5\height}{
	\setlength{\fboxsep}{0pt}%
	\setlength{\fboxrule}{.5pt}%
	\fbox{\includegraphics[width=.4\columnwidth]{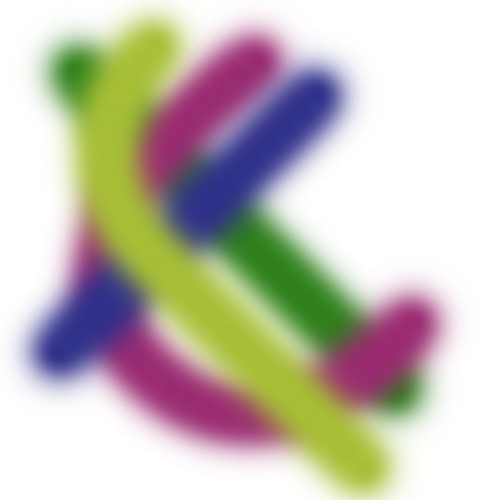}}
}
\raisebox{-0.5\height}{\includegraphics[width=.4\columnwidth]{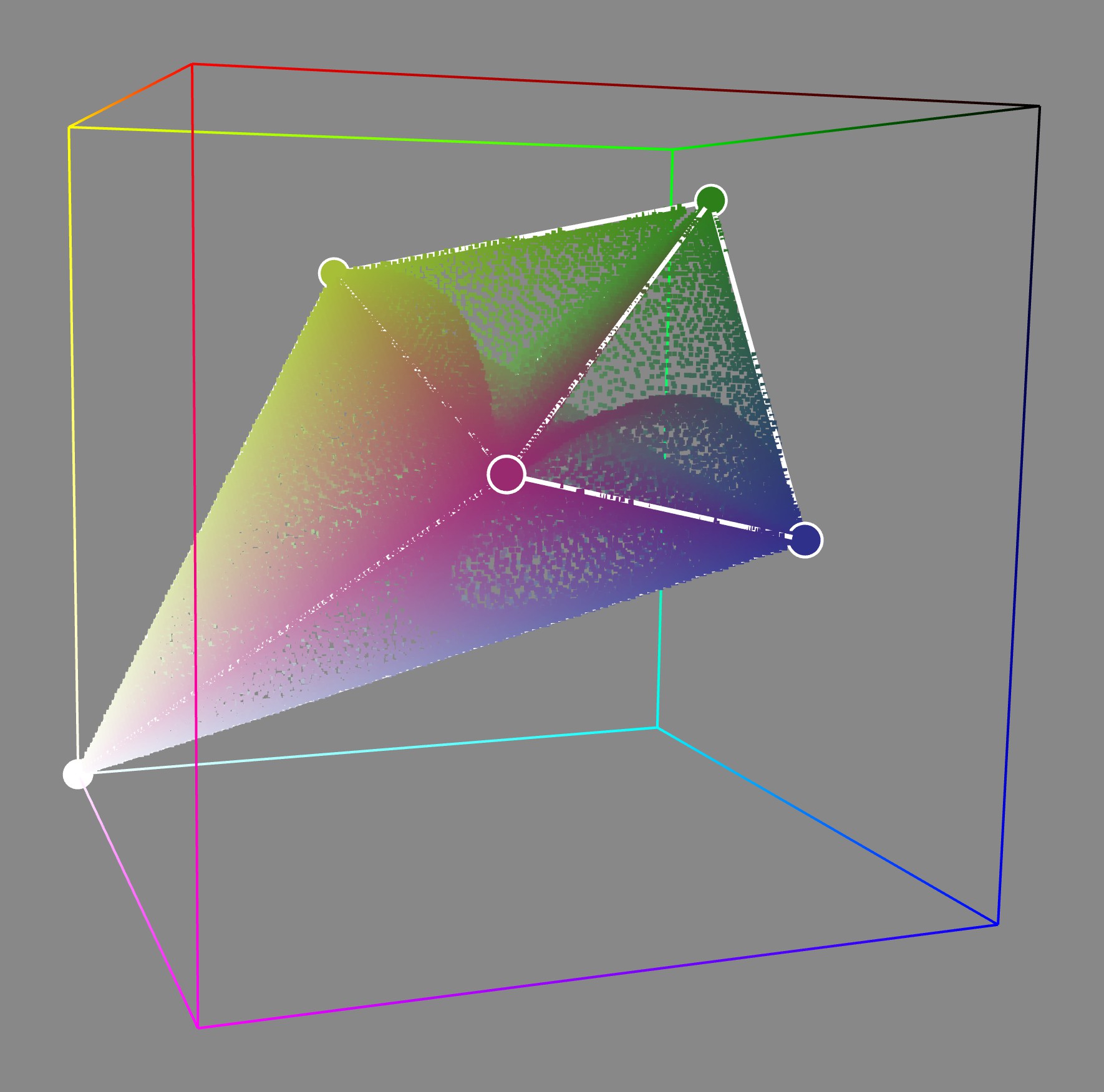}}
\caption{In RGB-space, the pixels of a digital painting (left)
lie in the convex hull of the original paint colors (right).
This is due to the linearity of the standard ``over'' blending operation
\protect\cite{Porter:1984:CDI}.
This linearity manifests as linear elements such as lines, faces, and solid polyhedra.
}
\label{fig:rgbspace}
\end{figure}

The first half of our pipeline identifies the colors used to paint the image (Section~\ref{sec:colors}).
All colors in the image must lie within the convex hull of these colors;
therefore we first compute the exact convex hull of the painting's pixels in RGB-space.
We then compute a simpler, approximate convex hull with user-tunable parameters.
The simplification is computed in two stages, first by plane fitting to hull faces, and then by clustering the resulting vertices.
These simplified vertices are the paint colors.

The second half of our pipeline takes as input a user-provided ordering of paint colors (RGB),
and computes the corresponding per-pixel opacity values to produce RGBA layers.
Each layer models a coat of paint.
Our computation solves a polynomial system of equations via energy minimization.
The polynomial equations express that the composition of all layers should reproduce the input image.
To solve this underconstrained problem, we introduce terms to maximize translucency and spatial coherence.

\section{Identifying Paint Colors}
\label{sec:colors}

The first step in our pipeline identifies the colors used to paint the image.
In a digital painting, many pixels will have been painted over
multiple times with different paint colors.
Because the paint compositing operation is a linear blend between two paint colors (Equation~\ref{eq:porterduff}),
all pixels in the painting lie
in the RGB-space convex hull
formed by the original paint colors.
Equivalently, any pixel color $\p$ can be expressed as the linear combination of the original paint colors $\c_i$:
$ \p = \sum w_i \c_i$ for some weights $w_i \in [0,1]$ with $\sum w_i = 1$.
Note that in this expression,
the $w_i$ are not opacity values. %
This property is true for Porter-Duff ``over'' compositing,
but not true for nonlinear compositing such as the Kubelka-Munk model of pigment mixing or layering \cite{Budsberg:2007:PCD}.
Figures~\ref{fig:teaser}, \ref{fig:rgbspace}, and \ref{fig:gallery}
display pairs of images and their pixels in RGB-space.

To identify the colors used to paint the digital image,
we first compute the RGB-space convex hull of all
observed pixel colors.
In practice,
if a paint color was always applied semi-transparently,
the convex hull will be overly complex (too many vertices).
This is because semi-transparent paint will not
produce any pixels with the paint color itself.
This manifests as ``cut corners'' or extra faces
in the convex hull.
To simplify the convex hull,
we uniformly sample points on the surface of the hull
and then perform plane fitting on the
samples.
We iteratively (a) find the best-fitting plane to the samples via RANSAC and (b) remove the associated samples. Iteration terminates when fewer than a small fraction of the samples remain (\emph{termination fraction} in Table~\ref{table:parameters}).
RANSAC requires a distance threshold parameter to determine how close a point must be to a plane to be considered an inlier; we use 3.0 for all our examples.
Finally, we recompute the convex region bounded
by the set of oriented planes \cite{Berg2008computational}.
The vertices of the convex region comprise the identified paint colors.
The oriented planes are positioned in space such that nearly all
points are inside the half-space and the vertices of the convex region are
within the RGB cube. The parameter which controls the planes' positions
is the fraction of all points inside the half-space
(\emph{inside fraction} in Table~\ref{table:parameters}).

We involve the user in the simplification process,
as it is difficult to find a set of parameters that provide satisfactory results in all cases.
We allow the user to choose RANSAC parameters.
We also allow the user to further simplify the
identified paint colors by choosing the bandwidth of
a mean-shift~\cite{MeanShift-2002} clustering operation
performed on the colors or by directly removing some colors.
``Over'' color compositing, while linear, is not commutative.
For $n$ layers, there are $n!$ orderings.
Because of the large possibility space, and the unknown
semantics of the colors,
we do not automate the determination of the layer order.
In our experiments, we computed opacity values for all $n!$ layer orders (Section~\ref{sec:opacity})
and attempted to find automatic sorting criteria.
We experimented with the total opacity, gradient of opacity, and Laplacian of opacity,
but none matched human preference.
As a result, we require the user to choose the layer order for the extracted colors.
The next stage of our algorithm determines the
per-pixel opacity of each layer.

\begin{figure}
\centering
\includegraphics[width=.49\columnwidth]{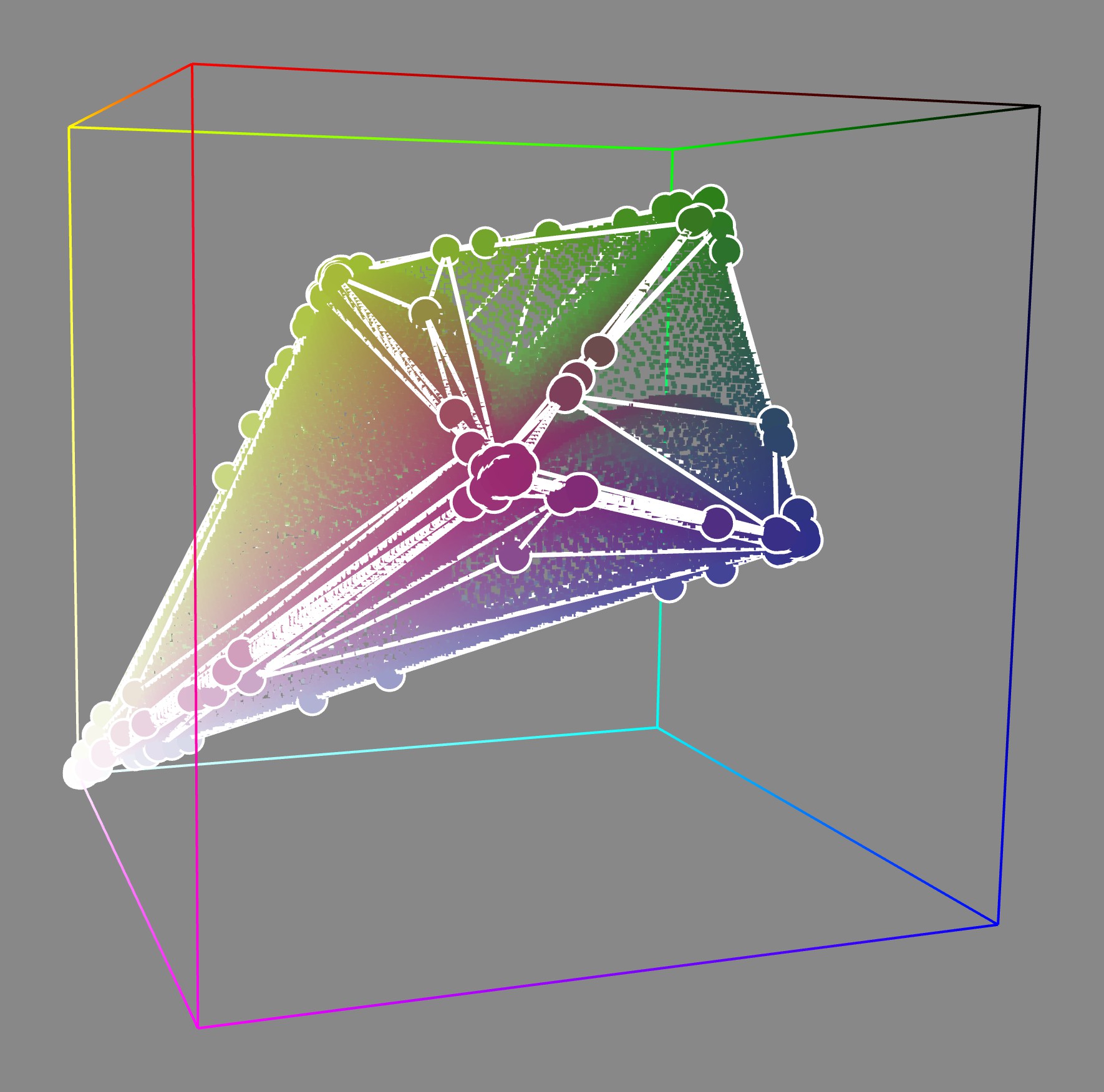}
\includegraphics[width=.49\columnwidth]{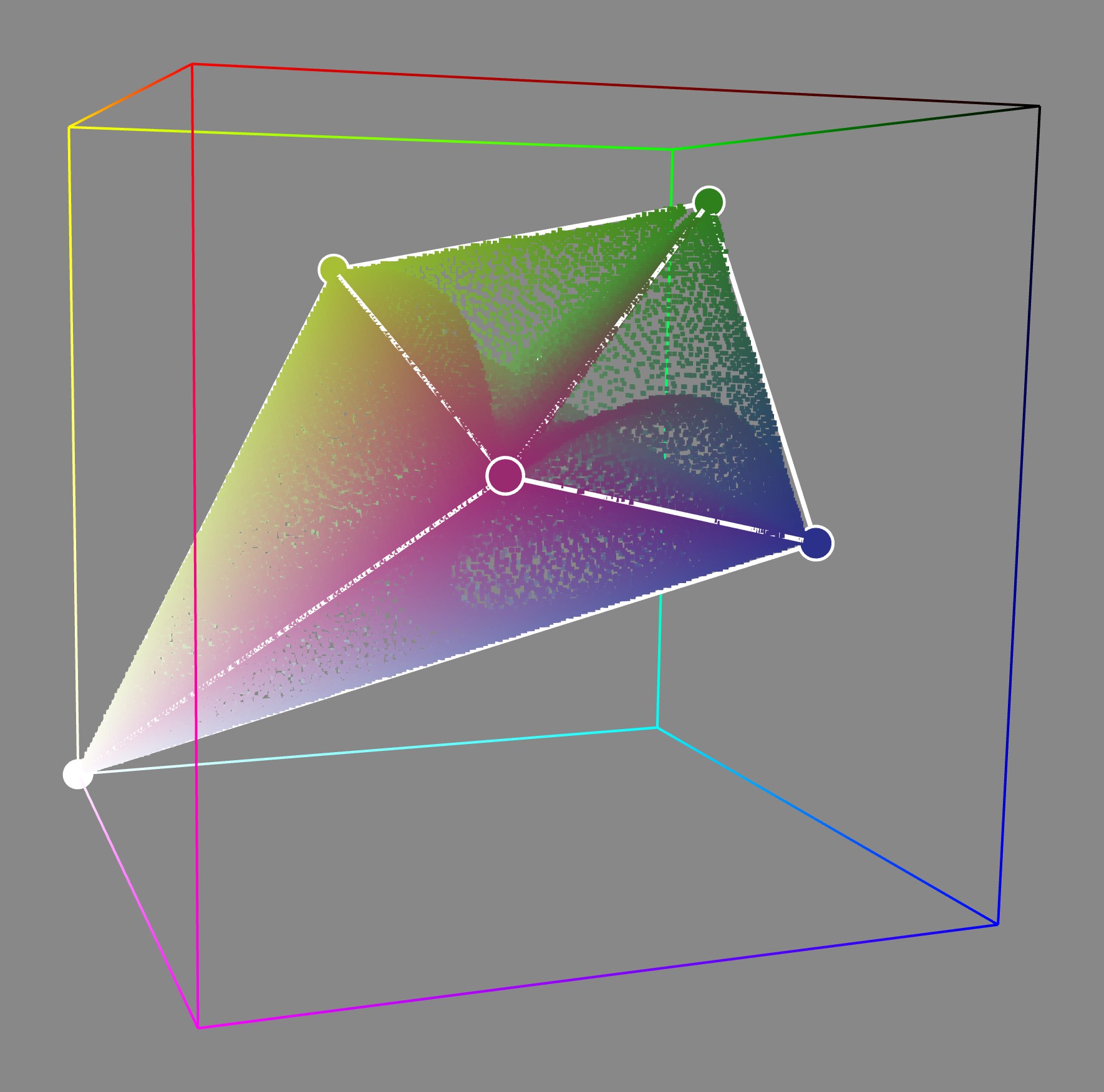}
\caption{The original convex hull (left) and the simplified hull (right).}
\label{fig:hull}
\end{figure}

\section{Determining Layer Opacity}
\label{sec:opacity}

\begin{figure}
\centering
\includegraphics[width=.5\columnwidth]{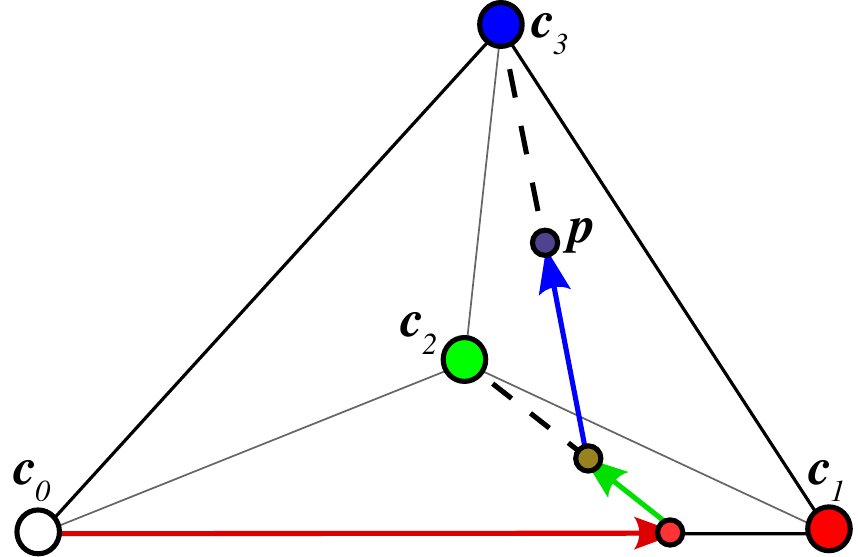}
\caption{A pixel can be represented as a path from the background color $\c_0$
towards each of the other colors $\c_1, \c_2, \c_3$ in turn.
The opacity values $\alpha_i$ determine the length of each arrow as a fraction
of the entire line segment to $\c_i$.}
\label{fig:blendpath}
\end{figure}

The final stage of our algorithm
computes opacity values for each layer,
at each pixel.
This stage takes as input
globally ordered layer colors.
Let $\{\c_i\}$ be the global, ordered sequence of RGB
layer colors,
and let $\c_0$ be the opaque background color.
Then, at each pixel, the observed color $\p$
can be expressed as the recursive
application of ``over'' compositing (Equation~\ref{eq:porterduff}):
\begin{align}
\label{eq:poly_constraint}
	\p = \c_n + \sum_{i=1}^n \left[ (\c_{i-1} - \c_i) \prod_{j=i}^n (1-\alpha_j) \right]
\end{align}
where $\alpha_i$ is the opacity of $\c_i$.
Since colors $\p$ and $\c_i$ are three dimensional (RGB),
this is a system of 3 polynomial equations with \#layer unknowns.
(For translucent RGBA input images,
premultiplied colors should be used.
Equation~\ref{eq:poly_constraint} becomes a system of
4 polynomial equations.
The background layer $\c_0$ becomes transparent---the zero vector---while the
remaining global layers colors $\c_i$ are treated as opaque.)

There will always be at least one solution,
because $\p$ lies within the convex hull of the $\c_i$.
When the number of layers is less than or equal to 3
(excluding the translucent background in case of RGBA),
there is, in general, a unique solution.
It can
be obtained directly by projecting $\p$
along the line from the top-most layer color $\c_n$
onto the simplex formed by $\c_0 \hdots \c_{n-1}$,
and so on recursively (Figure~\ref{fig:blendpath}).
However, if a layer is opaque (other than the bottom layer)
or the number of layers is greater than three,
there are infinitely many solutions.\footnote{If a solution with at most three
non-transparent layers \emph{per-pixel} is desired,
one can choose, for each pixel, any tetrahedron using convex hull vertices that encloses it.
For example, one may choose for a pixel the tetrahedron with smallest total alpha.}
(For numerical reasons, it is problematic if a layer
is nearly opaque---a situation which arises quite often.)

\paragraph{Regularization}
To choose among the solutions, we introduce two
regularization terms.
Our first regularization term penalizes opacity;
absent additional information, a completely occluded layer
should be transparent.
\begin{align}
\label{eq:energy_opaque}
    E_{\mathit{opaque}} = \sum_{i=1}^n -(1-\alpha_i)^2
\end{align}
Note that we do not naively minimize $\alpha^2$,
because (intuitively) $1 = 0^2 + 1^2 > \epsilon^2 + (1-\epsilon)^2 = 1 - 2\epsilon (1 - \epsilon)$.
In words, the squared opacity decreases when reducing
the opacity of an opaque layer and increasing the opacity
of a transparent layer.
\begin{wrapfigure}[8]{r}{1in}
\vspace{-1em}
\includegraphics[width=\linewidth]{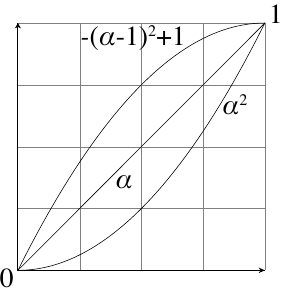}
\end{wrapfigure}
With our expression, the situation is reversed: $-1 = -(1-0)^2  - (1-1)^2  < -(1-\epsilon)^2  - \left(1-(1-\epsilon)\right)^2  = -1 + 2\epsilon$. Our energy decreases when reducing the opacity of a nearly transparent layer and increasing the opacity of a nearly opaque layer.
As a result, we obtain a sparser solution
(see inset).

For spatial continuity,
a local minimum of $E_{\mathit{opaque}}$ alone
is sometimes preferable to the global minimum.
Our second regularization term is the Laplacian energy of the $\alpha_i$ values.
\begin{align}
\label{eq:energy_spatial}
    E_{\mathit{spatial}} = \sum_{i=1}^n (\Delta \alpha_i)^2
\end{align}
where $\Delta \alpha_i$ is the Laplace operator,
or difference between the pixel's $\alpha_i$ and the average of its neighboring pixels' $\alpha_i$'s.
(We have obtained similar and sometimes better results with the Dirichlet energy, $\sum (\nabla \alpha_i)^2$.)

We minimize these two terms subject to the polynomial constraints (Equation~\ref{eq:poly_constraint}) and $\alpha_i \in [0,1]$.
We implement the polynomial constraints as a least-squares penalty term $E_{\mathit{polynomial}}$:
\footnote{The units of the three $E_{\mathit{polynomial}}$ expressions are squared color differences. Our RGB colors lie within $[0,255]$, so the implied weight is $255^2$.}
\begin{align}
\label{eq:energy_polynomial}
    E_{\mathit{polynomial}} = \left( \c_n - \p + \sum_{i=1}^n \left[ (\c_{i-1} - \c_i) \prod_{j=i}^n (1-\alpha_j) \right] \right)^2
\end{align}
The combined energy expression that we minimize is:
\begin{align*}
    E_{\mathit{polynomial}} + w_{\mathit{opaque}} E_{\mathit{opaque}} + w_{\mathit{spatial}} E_{\mathit{spatial}}
\end{align*}

\begin{figure}
\includegraphics[width=\columnwidth]{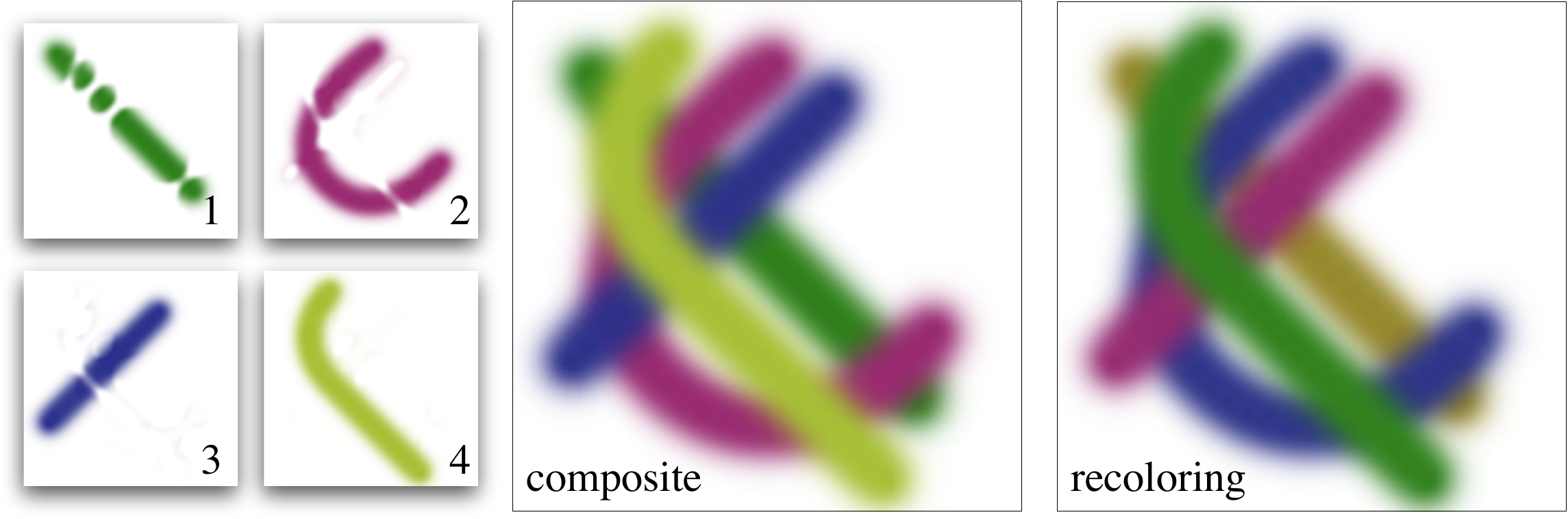}
\caption{Decomposed layers, their composition, and subsequent recoloring
from the example in Figures~\ref{fig:rgbspace} and \ref{fig:hull}.}
\label{fig:opacity_example}
\end{figure}

For the example shown in Figures~\ref{fig:rgbspace} and \ref{fig:hull},
the recovered layers and reconstruction are shown in Figure~\ref{fig:opacity_example}.

\section{Results}
\label{sec:results}

\begin{figure*}
\centering
\includegraphics[height=9.2in]{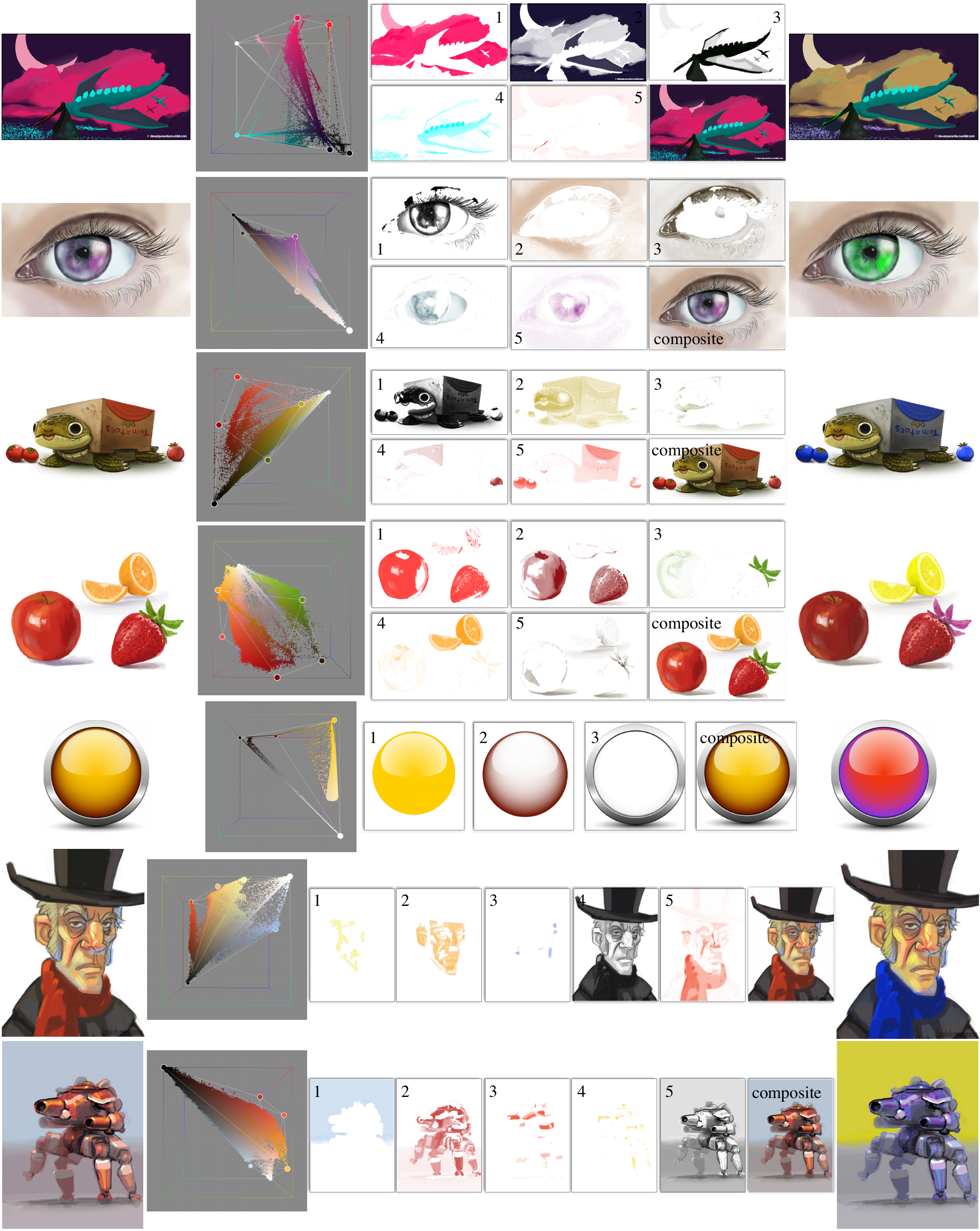}
\caption{Digital paintings, their RGB-space simplified convex hulls, decomposed layers, recomposition, and re-edits.
From top to bottom, original artwork \copyright\ 
Karl Northfell;
reddit user TwitchHD;
Piper Thibodeau;
Ranivius;
Roman Sotola; %
Dani Jones;
Adam Saltsman.} %
\label{fig:gallery}
\end{figure*}

Figures~\ref{fig:teaser} and \ref{fig:gallery} show the decomposition
of a variety of digital paintings, their recomposition, and layer-based edits.
The decomposed layers reproduce the input image without visually perceptible differences.
This is because the approximate convex hulls cover almost every pixel in RGB-space (Section~\ref{sec:colors}),
and the polynomial constraints in the energy minimization ensure that satisfying opacity values are chosen (Section~\ref{sec:opacity}).
See Table~\ref{table:parameters} for parameters.

These decomposed layer representations make it straightforward
to isolate edits such as recoloring a part of the image.
We have found parameter tuning to obtain a simplified convex hull to be straightforward.
Choosing the layer order and optimization weights is more time consuming,
as it can be difficult to predict which parameters will produce a meaningful set of layers.
For example, when the set of colors includes shades of red and black,
the boundary between the red layers may not match user expectations.
Nevertheless, we have been able to obtain useful decompositions for a large variety of examples.

\paragraph{Implementation}
We use QHull~\cite{QHull} for convex hull computation,
PCL~\cite{PCL} for RANSAC plane fitting,
and \textsc{L-BFGS-B}~\cite{L-BFGS-B} for optimization.
Our algorithms were written in Python
and vectorized using NumPy/SciPy.
Our implementation is not multi-threaded.

To improve the convergence speed of the numerical optimization,
we minimize our energy on recursively downsampled images
and use the upsampled solution as the initial guess
for the larger images (and, eventually, the original image).
We down/upsampled by factors of two.
We used $\alpha_i=0.5$ as the initial guess for the smallest image.
This progressive optimization reduced running time by a factor of four.

\paragraph{Performance}
Paint color identification (Section~\ref{sec:colors})
takes a few seconds to compute the simplified convex hull;
the bottleneck is the user choosing the desired
amount of simplification.
Computing layer opacity (Section~\ref{sec:opacity}) entails solving a nonlinear optimization procedure.
As we implemented our optimization in a multi-resolution
manner, the user is able to quickly see a preview of the result (e.g.\ less than 20 seconds for a $100\times64$-pixel image).
This is important for experimenting with different layer orders and energy weights.
Larger images are computed progressively as the multi-resolution optimization converges on smaller images;
for a 500 or 1000 pixels wide image, the final optimization
can take anywhere from 20 minutes to an hour or longer to converge.

\begin{table}
\centering

\resizebox{\columnwidth}{!}{

\newcolumntype{H}{>{\setbox0=\hbox\bgroup}c<{\egroup}@{}}

\begin{tabular}{lllrrrrH}
\toprule
& \multicolumn{1}{c}{plane fitting} & \multicolumn{2}{>{\centering\arraybackslash\hspace{0pt}}b{3cm}}{convex hull reconstruction} & \multicolumn{2}{c}{optimization} \\
\cmidrule(r){2-2} \cmidrule(lr){3-4} \cmidrule(l){5-6}
name & \multicolumn{1}{>{\centering\arraybackslash\hspace{0pt}}b{1.5cm}}{termination fraction} & \multicolumn{1}{>{\centering\arraybackslash\hspace{0pt}}b{1cm}}{inside fraction} & \multicolumn{1}{>{\centering\arraybackslash\hspace{0pt}}b{1.5cm}}{mean-shift bandwidth} & $w_{\mathit{opaque}}$ & $w_{\mathit{spatial}}$ & image size & \#layers \\
\midrule
apple                     & 0.05 & 0.99 & 40 & 100 & 1000 & $500 \times 453$ & 5 \\
Figure~\ref{fig:rgbspace} & 0.01 & 0.99 & 6 & 20 & 800000 & $500\times500$ & 4 \\
bird                      & 0.01 & 0.999 & 15 & 200 & 10000 & $640\times360$ & 5 \\
eye                       & 0.1 & 0.999 & 10 & 100 & 1000 & $630\times380$ & 5 \\
turtle                    & 0.1 & 0.999 & 30 & 100 & 10000 & $525\times250$ & 5 \\
fruits                    & 0.05 & 0.999 & 40 & 100 & 1000 & $650\times414$ & 5 \\
light                     & 0.1 & 0.95 & 31 & 100 & 1000 & $504\times538$ & 3 \\
scrooge                   & 0.1 & 0.97 & 31 & 100 & 1000 & $410 \times 542$ & 5 \\
robot                     & 0.1 & 0.995 & 30 & 100 & 1000 & $450 \times 600$ & 5 \\
\bottomrule
\end{tabular}

}

\caption{Parameters used in our pipeline.}
\label{table:parameters}
\end{table}

\section{Conclusion}
\label{sec:conclusion}

The RGB-space of a digital painting
contains a ``hidden'' geometric structure.
Namely, the convex hull of this structure identifies
the original paint colors.
Given a set of colors and an order,
our constrained optimization
decomposes the painting into
a useful set of translucent layers.

\paragraph{Limitations}
Our technique has several notable limitations.
First, selecting per-pixel layer opacity values is, in general, an underconstrained problem.
Our optimization employs two regularization terms
to bias the result towards translucent and spatially coherent solutions.
However, this still may not match user expectations.
Second, we expect a global order for layers.
We use layers to represent the application of a coat of paint.
However, in the true editing history, a single color may have been applied multiple
times in an interleaved order with the other colors.
Third, hidden layer colors---those that lie within
the convex hull---cannot be detected by our technique.
We also do not allow colors to change during optimization;
we experimented with an energy term allowing layers colors to change
but found it difficult to control.
A related problem is paintings of e.g.\ rainbows;
when the convex hull encompasses all or much of RGB-space,
layer colors become uninformative (e.g.\ pure red, green, blue, cyan, magenta, yellow, and black).
Fourth, we require user input to choose the degree of simplification for the convex
hull, to choose the layer order, and to choose optimization weights.
While the number of layer orderings is factorial in the number of layers,
heuristics may exist to narrow the search space.
Finally, nonlinear color-space transformations, such as gamma correction,
distort the polyhedron. We ignore gamma information stored in input images.

\paragraph{Future Work}
In the future,
we plan to study the effects of nonlinear digital edits in RGB-space,
and to consider non-linear color compositing operations,
such as the Kubelka-Munk mixing and layering equations \cite{Tan:2015:DTL}.
This would allow us to decompose scans of physical paintings
and to support a larger variety of digitally created images.
We also plan to explore heuristics for automatically choosing a layer order
and degree of simplification to apply to the convex hull.
Although we focus on digital paintings,
the colors of any image (physical painting or photograph)
can still be covered by a convex hull in RGB-space,
and therefore decomposed into ``paint'' layers with our approach.
Finally, we plan to apply our per-pixel layer opacity values
towards segmentation; layer translucency is a higher-dimensional and more meaningful
feature than composited RGB color.

\section*{Acknowledgements}
We are grateful to Neil Epstein for an interesting conversation
about the algebraic structure of the solutions
to the multi-layer polynomial system.
This work was supported in part
by the United States National Science Foundation
(IIS-1451198, IIS-1453018, IIS-096053, CNS-1205260, EFRI-1240459, and AFOSR FA9550-12-1-0238)
and a Google research award.
Several experiments were run on ARGO,
a research computing cluster provided
by the Office of Research Computing
at George Mason University, VA. (URL:\url{http://orc.gmu.edu}).

\nocite{Tange2011a}

\bibliographystyle{acmsiggraph}
\bibliography{bib/crowdcreate,bib/timemap,bib/singlelayer}
\end{document}